\documentclass[a4paper,11pt]{article}
\usepackage{pos}
\usepackage{hyperref,setspace,subcaption}
\usepackage{mathtools}
\usepackage{graphicx}
\usepackage{epsfig}
\usepackage{bm}

\newcommand{\postscript}[2]{\setlength{\epsfxsize}{#2\hsize}
   \centerline{\epsfbox{#1}}}

\title{Gauging the cosmic ray muon puzzle with the Forward Physics Facility }
 \ShortTitle{Gauging the cosmic ray  muon puzzle with the FPF}

\author*[ab]{Sergio J. Sciutto}
\author[cde]{Luis A. Anchordoqui}
\author[ab]{Carlos Garc\'{\i}a Canal}
\author[f]{Felix Kling}
\author[c]{Jorge F. Soriano}
\onbehalf{for the FPF Initiative}
\affiliation[a]{Instituto de F\'{\i}sica La Plata - CONICET,  C.C. 69,
 (1900) La Plata, Argentina}
\affiliation[b]{Departamento de F\'{\i}sica, Universidad Nacional de La Plata, C.C. 69, (1900) La Plata, Argentina}
 \affiliation[c]{Department of Physics and Astronomy,  Lehman College, City University of
  New York, NY 10468, USA
}
\affiliation[d]{Department of Physics,
 Graduate Center, City University
  of New York,  NY 10016, USA
}
\affiliation[e]{Department of Astrophysics,
 American Museum of Natural History, NY
 10024, USA
}

\affiliation[f]{Deutsches Elektronen-Synchrotron DESY, Notkestr. 85, 22607 Hamburg, Germany}

\emailAdd{sciutto@fisica.unlp.edu.ar}

\abstract{We investigate the observed muon deficit in air shower simulations when compared to ultrahigh-energy cosmic ray (UHECR) data. Gleaned from the observed enhancement of strangeness production in ALICE data, the associated $\pi \leftrightarrow K$ swap is taken as a cornerstone to resolve the muon puzzle via its corresponding impact on the shower evolution. We develop a phenomenological model in terms of the $\pi \leftrightarrow K$ swapping probability $F_s$. We provide a  parametrization of $F_s (E^{\rm (proj)}, \eta)$ that can accommodate the UHECR data, where $E^{\rm (proj)}$ is the projectile energy and $\eta$ the pseudorapidity. We also explore a future game plan for model improvement using the colossal amount of data to be collected by LHC neutrino detectors at the Forward Physics Facility (FPF). We calculate the corresponding sensitivity to $F_s$ and show that the FPF experiments will be able to probe the model phase space.}

\ConferenceLogo{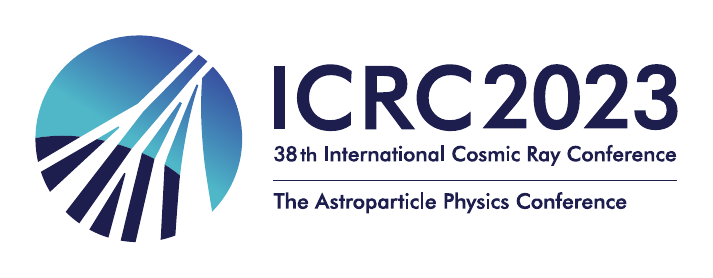}

\FullConference{%
38th International Cosmic Ray Conference (ICRC2023)\\
  26 July - 3 August, 2023\\
  Nagoya, Japan}

\begin{document}
\maketitle

\section{Introduction}

Besides addressing major questions in astrophysics, ultra-high-energy cosmic ray (UHECR) experiments provide unique access to scattering processes at center-of-mass energies well beyond those achieved in $pp$ collisions at the Large Hadron Collider (LHC), and wherefore deliver a priceless examination of particle interactions below the fermi distance~\cite{Anchordoqui:2018qom}. Along this line, UHECR experiments have been observing significant discrepancies between the number of observed muons in extensive air showers and model predictions~\cite{PierreAuger:2014ucz,PierreAuger:2016nfk,TelescopeArray:2018eph}. This observation is generally referred to as the UHECR muon puzzle~\cite{Albrecht:2021cxw}. The muons seen by UHECR experiments are of low energy (a few to tens of GeV). They are produced at the end of a cascade of hadronic interactions, where the dominant process is soft forward hadron production, which cannot be calculated from first principles in perturbative QCD. Instead, effective theories are used to describe these interactions. Detailed simulations~\cite{Ulrich:2010rg} have shown that the hadron multiplicity and, in particular, the hadron species at forward pseudorapidities of $\eta \gg 2$ have the largest impact on muon production in air showers.

A, seemingly different, but in fact closely related subject has been the
observation of a strangeness enhancement at mid-rapidities, i.e., at pseudorapidity regions of $2.8<\eta<5.1$ and $-3.7<\eta<-1.7$ by the ALICE collaboration~\cite{ALICE:2016fzo}. This observation indicates that the amount of forward strangeness production seems to be of particular relevance in order to understand the origin of the UHECR muon puzzle~\cite{Allen:2013hfa,Anchordoqui:2016oxy,Baur:2019cpv,Manshanden:2022hgf}. We have shown elsewhere~\cite{Anchordoqui:2019laz}  that  none of the hadronic models in the market correctly reproduce the main tendencies of ALICE data,
especially for the description of multi-strange hadron
production.  In this communication we show that if the muon puzzle is related to strangeness enhancement in the forward region, then external input to guide (hadronic) model builders could come from LHC data, particularly from new experiments at the Forward Physics
Facility (FPF)~\cite{Anchordoqui:2021ghd,Feng:2022inv}.

Very recently, FASER has observed the first neutrinos from LHC collisions~\cite{FASER:2023zcr}. During the  high-luminosity era, LHC collisions will provide an enormous flux of neutrinos and muons originating from the decay of light hadrons, such as pions and kaons. The ratio of charged kaons to pions, for which the ratio of electron and muon neutrino fluxes is a proxy, can be measured by FPF experiments~\cite{Kling:2021gos,Feng:2022inv,Anchordoqui:2021ghd}. Muon neutrino fluxes are a measurement of pions, whereas both muon and electron neutrinos are produced via kaon decay. However, $\nu_\mu$ and $\nu_e$ populate different energy and rapidity regions, which will allow to disentangle different neutrino origins to get an estimate of the pion to kaon ratio. In addition, neutrinos from pion decay are more concentrated around the line-of-sight than those of kaon origin, given that $m_\pi < m_K$, and thus neutrinos from pions  obtain less additional transverse momentum than those from kaon decays. Thereby, the closeness of the neutrinos to the line-of-sight, or equivalently their rapidity distribution, can be used to disentangle different neutrino origins to get an estimate of the pion to kaon ratio.

Very recently, we constructed a phenomenological (one-parameter) model based on the $\pi \Leftrightarrow K$ swapping probability that can accomodate the discrepancies between data and simulations~\cite{Anchordoqui:2022fpn}. In this communication, we first provide a model uptake by considering a
three-parameter model, in which the swapping probaility has dependence on the projectile energy and pseudorapidity. After that we explore possibilities for model improvement using the colossal amount of data to be collected by LHC neutrino detectors at the FPF. We calculate the corresponding sensitivity to the swapping probability  and show that the FPF experiments will be able to probe the model phase space.

\section{The piKswap (/{\tt p\=ekasw\"ap}/) model  }

We begin by discussing general aspects of our phenomenological model and  investigating how the model impacts the evolution of extensive air showers,  while addressing the muon puzzle. To describe the shower evolution we adopt the AIRES simulation engine~\cite{Sciutto:1999jh}, which provides full space-time particle propagation in a realistic environment.\footnote{For details on the AIRES version (19.04.08) adopted herein, see {\tt http://aires.fisica.unlp.edu.ar}.}
We developed a new module to account for the possible enhancement of strangeness production in high-energy hadronic collisions. Every time an hadronic collision is processed, the list of secondary particles obtained from an external event generator  (for our analysis we adopt \textsc{Sibyll-2.3d}~\cite{Riehn:2019jet}), is scanned by the new module before passing it to the main particle propagating engine. It is noteworthy that our results are not substantially dependent on the hadronic interaction model.

The main characteristics of the new AIRES module are stamped in the following parameteres: {\it (i)}~{\bf Swapping fraction}, $F_s(E^{\rm (proj)},\eta)$, which controls the number of secondary pions that are
     affected by the change of identity as a function of the projectile energy $E^{\rm (proj)}$
and the center of
     mass pseudorapidity of the secondary particles $\eta$. Note that  $0\le F_s \le 1$.
{\it (ii)}~{\bf Maximum swapping fraction}, $f_{s}^{(\rm max)}$, which determines the
     maximum  value of $F_s$. {\it (iii)}~{\bf Minimum projectile energy},  $E_{\rm pmin}$, which defines the low-energy threshold, i.e., particle swapping is performed in hadronic collisions whose
     projectile kinetic energy is larger than this energy. $E_{\rm
       pmin}$ must be larger than 900~MeV. {\it (iv)}~{\bf Reference projectile energy},  $E_{\rm pref}$, which is a
     parameter of $F_s(E_{\rm proj},\eta)$; see
     below. $E_{\rm pref}$ must be larger than $E_{\rm pmin}$. {\it (v)}~{\bf Minimum secondary energy},  $E_{\rm smin}$, which describes the minimum energy in the swapping process,
i.e., secondary particles with kinetic energies below $E_{\rm smin}$
     are always left unchanged.  $E_{\rm smin}$ must be larger than 600~MeV. {\it (vi)}~{\bf Minimum secondary pseudorapidity}, $\eta_{\rm min}$,  which indicates the lowest pseudorapidity for the swapping process, i.e,
     particle swapping is never performed in secondary particles whose
     pseudorapidity is in absolute value less or equal than this
     parameter (untouched ``central'' zone). $\eta_{\rm min}$ must be positive.
{\it (vii)}~{\bf Reference sec pseudorapidity}, $\eta_{\rm ref}$, which is a parameter of $F_s(E_{\rm proj},\eta)$; see
     below. $\eta_{\rm ref}$ must be larger than $\eta_{\rm min}$.

The logics for post-processing hadronic collisions is as follows. During shower simulation, hadronic collisions are processing via calls
to a hadronic interaction generator. The input parameters for these
calls are the projectile identity ($p_{\rm id}$) and its kinetic energy
($E^{\rm (proj)}$), and the target
identity. On return, the generator provides a list of $N_{\rm sec}$
particles, specifying their identity ($s_{{\rm id}i}$,
$i=1,\ldots,N_{\rm sec}$), energy ($E^{\rm (sec)}_i$,
$i=1,\ldots,N_{\rm sec}$), momentum, etc.

All the returned secondary particle lists undergo a post-pocessing
procedure, just before they are stacked into the particle stacks for
further propagation. The post-processing algorithm obeys the following
rules: {\it (i)}~If $E^{\rm (proj)}<E_{\rm pmin}$  then no
  action is taken; the secondary particle list remains
  unchanged. {\it (ii)}~If $E^{\rm (proj)}>E_{\rm pmin}$ then
\begin{equation}
    f_s^{\rm (coll)} = f_{s}^{\rm (max)} \times \left\{
         \begin{array}{lcl}{\displaystyle
            \frac{\log E^{\rm (proj)}-\log E_{\rm pmin}}{%
                  \log E_{\rm pref}-\log E_{\rm pmin}}}
            & \hbox{if} & E^{\rm (proj)} < E_{\rm pref} \\*[5mm]
            1 & \hbox{if} & E^{\rm (proj)} \ge
            E_{\rm pref}
         \end{array} \right.
\end{equation}
is evaluated.  In this way, the effective $f_s$ parameter for the
current collisions ramps up from zero to a maximum value when the
projectile energy goes from the minimum $E_{\rm pmin}$
to the ``reference'' value $E_{\rm pref}$, and remains in that maximum
value for projectile energies larger than $E_{\rm pref}$. {\it (iii)} The list of secondaries is scanned
  and processed as follows: {\it (1)}~All the secondary pions whose kinetic energies (lab system) are
  larger than $E_{\rm smin}$ are considered for
  identity swapping. Each of them is randomly selected with
  probability
\begin{equation}
 F_s(E^{\rm(proj)},\eta) =\left\{
\begin{array}{lcl}
0 &\hbox{if}& |\eta|\le \eta_{\rm min} \\*[4mm]
f_s^{\rm (coll)} \left[ {\displaystyle
    \frac{|\eta|-\eta_{\rm min}}{\eta_{\rm ref}-\eta_{\rm min}}}
    \right]
 &\hbox{if}& \eta_{\rm min} < |\eta| < \eta_{\rm ref} \\*[6mm]
f_s^{\rm (coll)} &\hbox{if}& |\eta|\ge \eta_{\rm ref}
\end{array}
\right.
\end{equation}
  where $\eta$ is the
  centre of mass pseudorapidity of the secondary. Note that
    secondaries within the
    ``central'' zone ($|\eta|\le \eta_{\rm min}$) are always
    left unchanged. {\it (2)}~In case of positive
  selection in the previous step,
 the identity is changed with the following criterion: {\it (a)}~Each $\pi^0$ is transformed onto $K^0_S$ of $K^0_L$, each case
  with 50\% probability; {\it (b)} Each $\pi^+$ ($\pi^-$) is transformed onto $K^+$ ($K^-$). {\it (3)}~Every time a secondary particle undergoes change of identity,
  the kinetic energy of the transformed particle is set via
 $E_{\rm new} = E_{\rm old} + m_{\rm old} - m_{\rm new}$, where
  ``old'' (``new'') refers to the selected secondary particle before
  (after) swapping, and $m$ stands for the rest energy of the
  corresponding particle.

\begin{figure}[htpb!]
  \begin{minipage}[t]{0.32\textwidth}
  \postscript{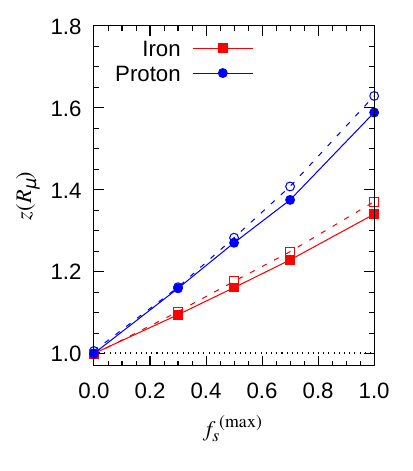}{0.99}
  \end{minipage}
  \begin{minipage}[t]{0.32\textwidth}
  \postscript{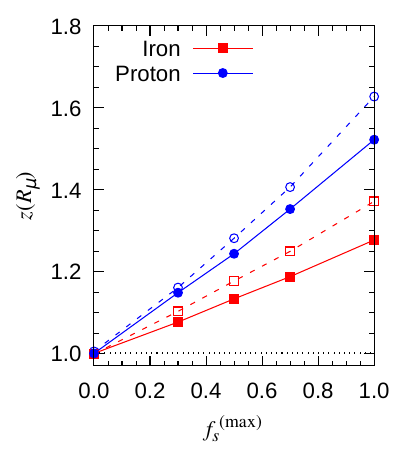}{0.99}
  \end{minipage}
  \begin{minipage}[t]{0.32\textwidth}
  \postscript{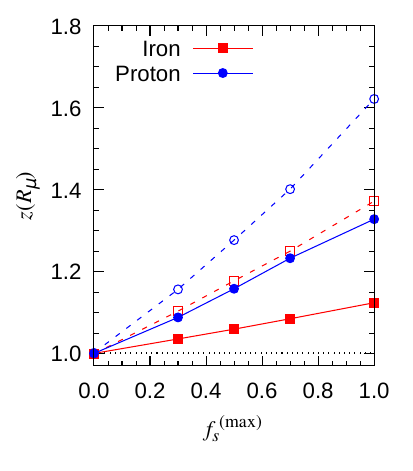}{0.99}
  \end{minipage}
  \caption{$z(R_\mu)$ versus $f_s^{\rm (max)}$, as obtained from simulations using AIRES+\textsc{Sibyll-2.3d} with the multi-parameter $\pi\leftrightarrow K$ swap model (solid lines).
 From left to right:
 $E_{\rm pref}= E_{\rm pmin}$, $\eta_{\rm ref} = 7$;
  $E_{\rm pref}=10^{19}$~eV, $\eta_{\rm ref} =\eta_{\rm min}$;
 $E_{\rm pref}= 10^{19}$ eV, $\eta_{\rm ref} = 7$.
 For comparison, we include in every plot the results corresponding to the basic one parameter model \cite{Anchordoqui:2022fpn} (dashed lines). All the simulations were done using $E_{\rm pmin}=10^{15}$~eV, $E_{\rm smin}=1$~GeV, and $\eta_{min} = 3$.
 \label{fig:zRmuVersusfsmax}}
\end{figure}

\begin{figure}[htb!]
 \begin{minipage}{0.49\textwidth}
\postscript{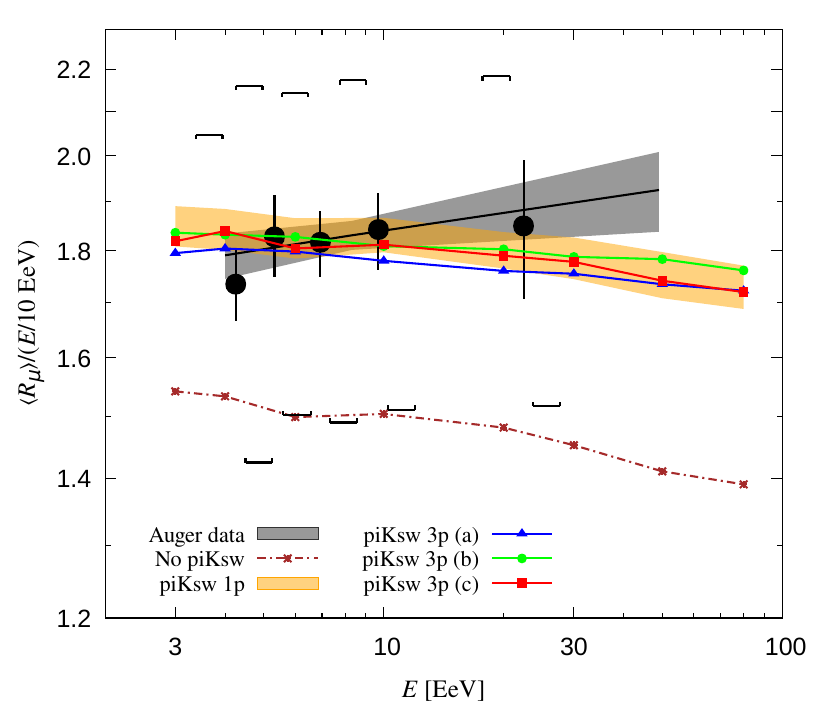}{1.}
 \end{minipage}
 \begin{minipage}{0.49\textwidth}
  \caption{Estimations of $R_\mu$ from AIRES+\textsc{Sibyll-2.3d} simulations superimposed over Auger data with statistical
\mbox{($\hspace{0.1em}\bullet\hspace*{-.66em}\mid\hspace{0.16em}$)}
 and systematic (\protect\rotatebox{90}{\hspace{-.075cm}[ ]})
 uncertainties~\cite{PierreAuger:2014ucz}, in the following cases: {\bf No piKsw}, no $\pi\leftrightarrow K$ swapping; {\bf piKsw 1p}, $\pi\Leftrightarrow K$ swapping using the one-parameter ($f_s$) model~\cite{Anchordoqui:2022fpn}, the shadowed area corresponding to $0.4\lesssim f_s\lesssim 0.5$; 
 {\bf piKsw 3p (a)}, the three-parameter model with  $E_{\rm pref}=10^{19}$~eV, $f_s^{\rm (max)}=0.8$, and $f_s=f_s^{\rm (coll)}$ for all secondary pions with $|\eta|> 3$ ($\eta_{\rm ref} =\eta_{\rm min}$);
 {\bf piKsw 3p (b)}, the three-parameter model with  $E_{\rm pref}=10^{19}$~eV, $f_s^{\rm (max)}=1.0$, and $\eta_{\rm ref} = 7$;
 {\bf  piKsw 3p (c)}, the three-parameter model with  $E_{\rm pref}=E_{\rm pmin}$, $f_s^{\rm (max)}=0.5$ for all $E^{\rm proj} \ge E_{\rm pmin}$ ($E_{\rm pref}= E_{\rm pmin}$), and $\eta_{\rm ref} = 7$. All the {\bf piKsw 3p} simulations were done using $E_{\rm pmin}=10^{15}$ eV, $E_{\rm smin}=1$~GeV, and $\eta_{min} = 3$. In all cases we adopted the mixed
 baryonic composition employed in~\cite{Anchordoqui:2022fpn}.
\label{fig:RmuVersusEprim}}
\end{minipage}
\end{figure}

 Using the output of AIRES simulations we first calculate the dimensionless muon content $R_\mu = N_{\mu}/N_{\mu,19}$  and then we determine the ratio
 \begin{equation}
z(R_\mu) = \frac{R_\mu(\hbox{piKswap On})}{R_\mu(\hbox{piKswap Off})} \,,
\end{equation}
where $N_\mu$ is the total number of muons (with $E_\mu > 300~{\rm MeV}$) at ground level and $N_{\mu ,19} = 1.455 \times  10^7$ is the average number of muons in simulated proton showers at $10^{19}~{\rm eV}$ with incident angle of $67^\circ$. 

In Fig.~\ref{fig:zRmuVersusfsmax} we show a comparison of $z(R_\mu)$ as predicted by the one-parameter model and  three different configurations of the multi-parameter model. In the left panel we consider a ramp in pseudorapidity but without a ramp in energy, in the middle panel a ramp in energy without a ramp in pseudorapidity, and in the right panel 
a ramp in both energy and pseudorapidty.  We can see that the one-parameter model is the one that predicts the largest increase in the number of muons. The plot on the right corresponds to the case where $f_s^{\rm (coll)} = f_s^{\rm (max)}$ is independent of $E^{\rm (proj)}$ for all projectile energies above the swapping threshold, and $f_s$ grows linearly with the pseudorapidity of the secondary pions until reaching a maximum for $|\eta|> 7$. In this case, no notable differences are found with the case of the one-parameter model~\cite{Anchordoqui:2022fpn}, probably indicating that the secondary kaons with maximum pseudorapidity are the ones that have the greatest impact in increasing the final number of muons. This effect is not appreciable when $f_s^{\rm (coll)}$ is allowed to vary with the energy of the projectile, but taking $f_s = f_s^{\rm (coll)}$ for all  $|\eta|>3$ (middle graph). In this case, a significant decrease in $z(R_\mu)$ is observed when compared to the model with constant $f_s$. The model featuring a ramp in both energy and pseudorapidity is the one that  predicts the smallest increase in the number of muons, and it actually requires $f_s \to 1$ to accommodate the data.

In the spirit of~\cite{Sciutto:2019pqs}, in Fig.~\ref{fig:RmuVersusEprim} we incorporate the change of the nuclear composition of the cosmic ray primary and analyze the variation of $\langle R_\mu \rangle/(E/10~{\rm EeV})$. The simple observation of Fig.~\ref{fig:RmuVersusEprim} allows one to clearly conclude that the originally proposed one-parameter  piKswap model shows a clear effectiveness in producing the required amount of muons to accommodate the data by simply introducing a $\pi \leftrightarrow K$ swapping regulated by $f_s$ in the range $0.4 \lesssim f_s \lesssim 0.5$. In fact, the  model uptake that includes possible variations of the swapping probability in terms of the energy and the pseudorapidity does not improve the prediction of the muon production. The model uptake introduced herein is not more efficient in doing the job and requires the introduction of other free parameters in the play. 

\section{A hard jigsaw puzzle to challenge the FPF experiments}

\begin{figure}[htb!]
  \begin{minipage}[t]{0.48\textwidth}
  \postscript{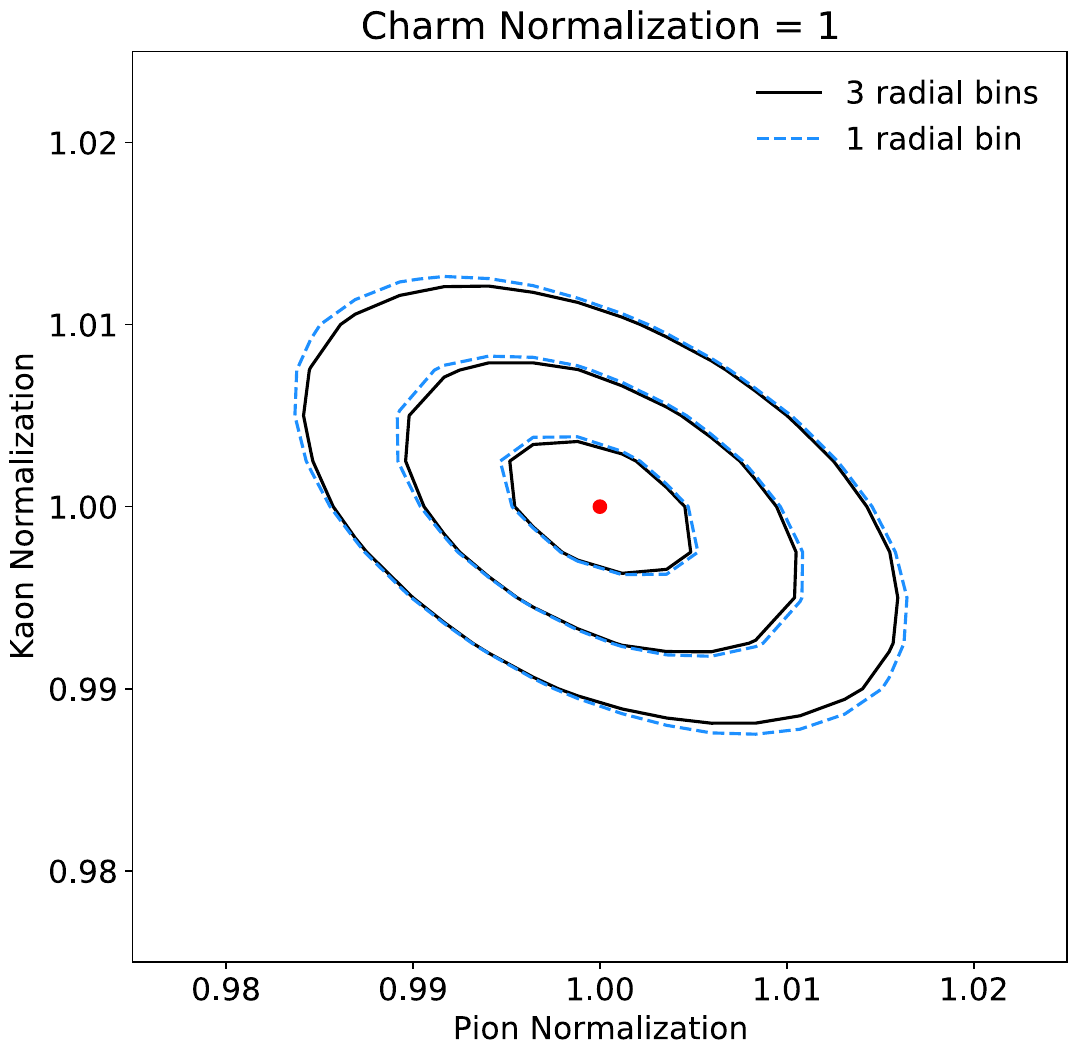}{0.95}
  \end{minipage}
\begin{minipage}[t]{0.48\textwidth}
    \postscript{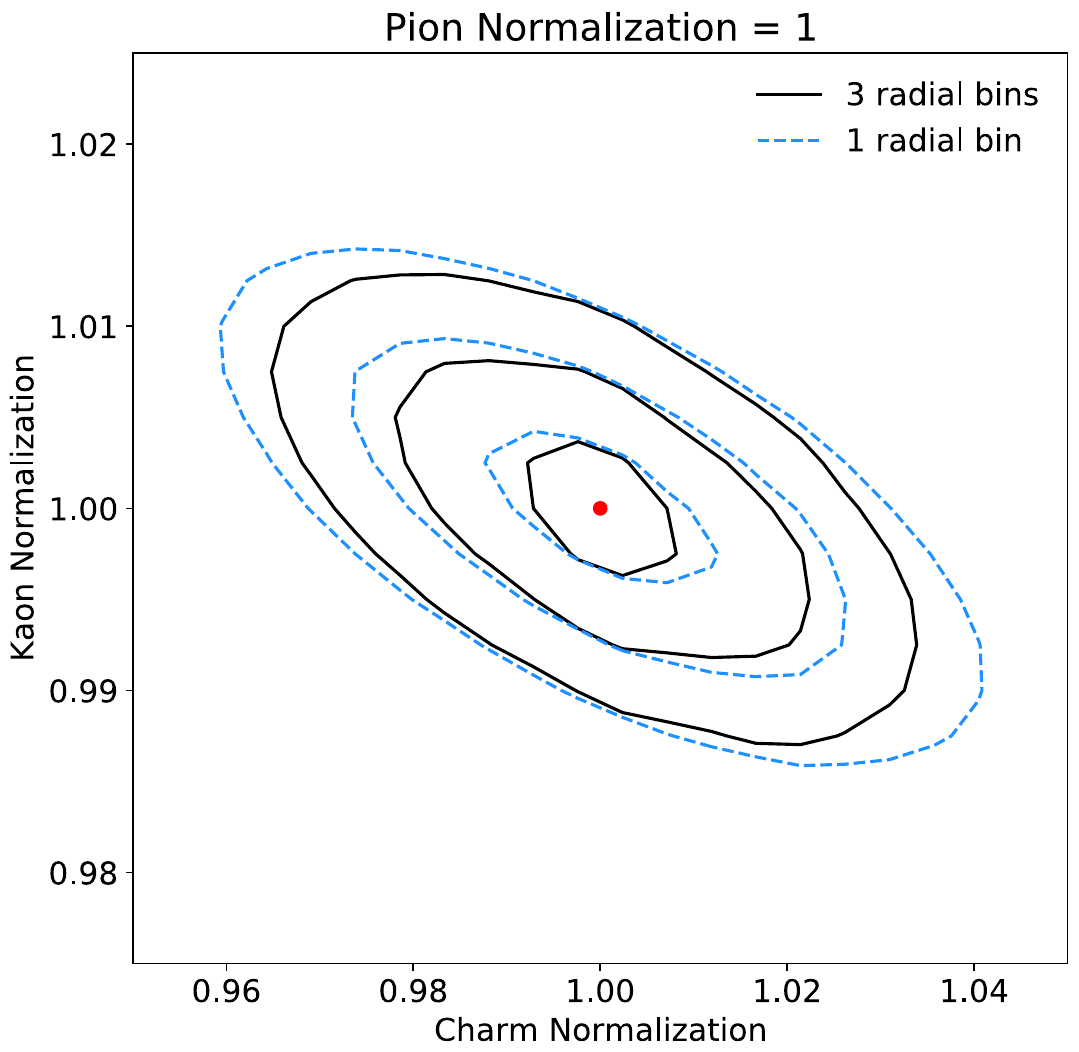}{0.95}
  \end{minipage}
\caption{Normalization of the pion, kaon, and prompt flux component obtained from a fit to the expected FLArE data based on \textsc{Sibyll-2.3d}~\cite{Riehn:2019jet} simulations. The lines indicate the $1\sigma$, $2\sigma$, and $3\sigma$ contours, respectively (taking into account only statistical uncertainties). The fit is applied using an analysis without radial binning (1 radial bin, dashed lines) and with radial binning (3 radial bins, solid lines).\label{fig:fits}}
\end{figure}

We now turn to answer the question of how well the FPF experiments could test hadronic models of strangeness  enhancement that can resolve the muon puzzle. Bearing this in mind, we first investigate how well the FPF experiments
could constrain forward particle production using neutrino measurements.
We adopt as a working hypothesis the neutrino fluxes resulting from numerical simulations of \textsc{Sibyll-2.3d}~\cite{Riehn:2019jet} with $\eta =4$.\footnote{We note in passing that there is an almost  negligible variation in the neutrino flux while changing the pseudorapidity cut from $\eta \geq 4$ to $\eta \geq 8$~\cite{Anchordoqui:2022ivb}.} Herein we refer to this as the baseline model. We introduce three free parameters, corresponding to the normalization of the pion, kaon, and charm flux component, which are used to fit the expected data. Our results are normalized using the design specifications of the FLArE detector, which is assumed to have a $1~{\rm m} \times 1~{\rm m}$ cross sectional area and a 10~ton target mass~\cite{Anchordoqui:2021ghd,Feng:2022inv}. The results of our analysis are encapsulated in Fig.~\ref{fig:fits}, which shows the $1\sigma$, $2\sigma$, and $3\sigma$ contours for two assumptions of the radial binning (taking into account only statistical uncertainties). The results shown in Fig.~\ref{fig:fits} indicate that FLArE data will provide strong constraints on the normalization of pion and kaon fluxes at the sub-percent level.

\begin{figure}[htb!]
  \postscript{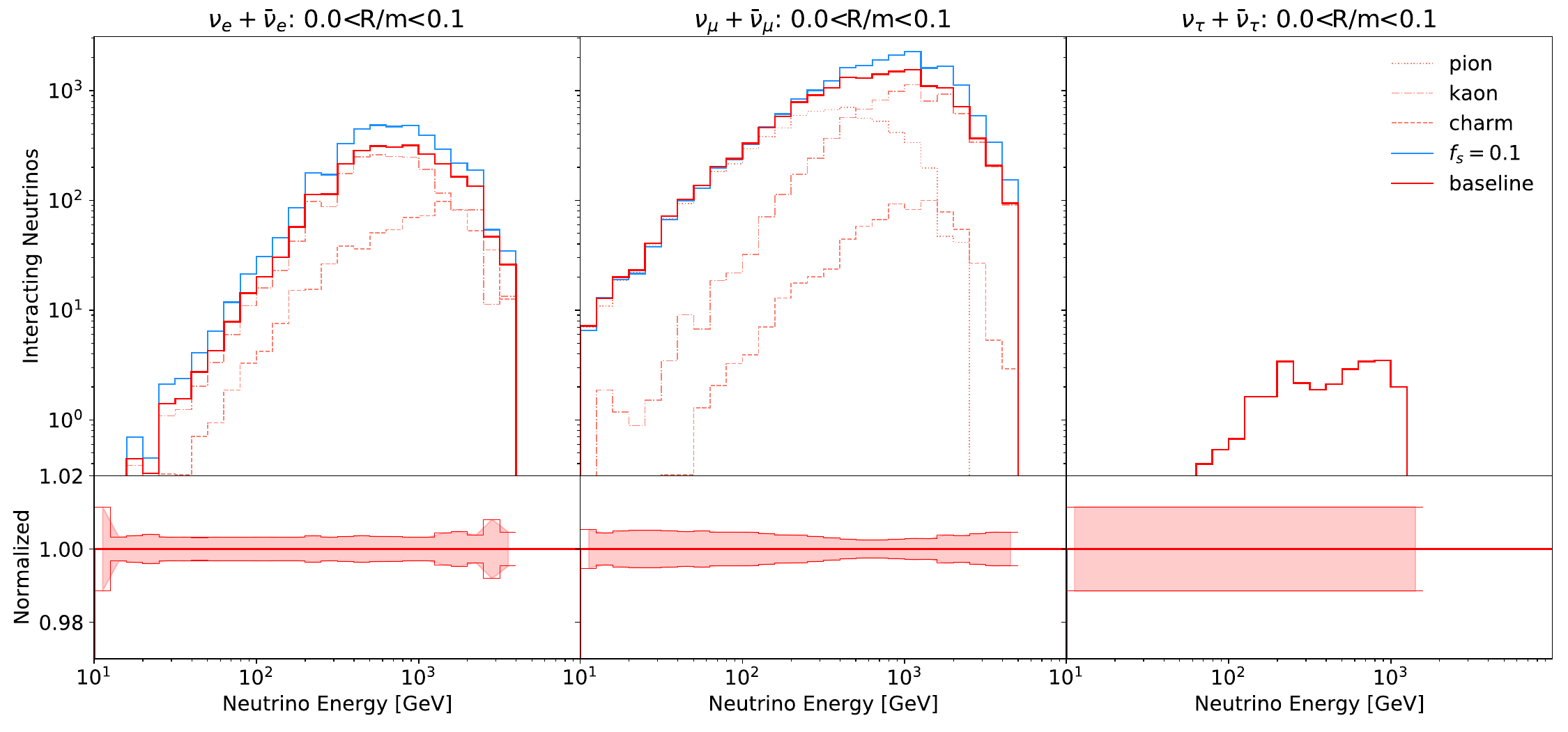}{0.8}
  \postscript{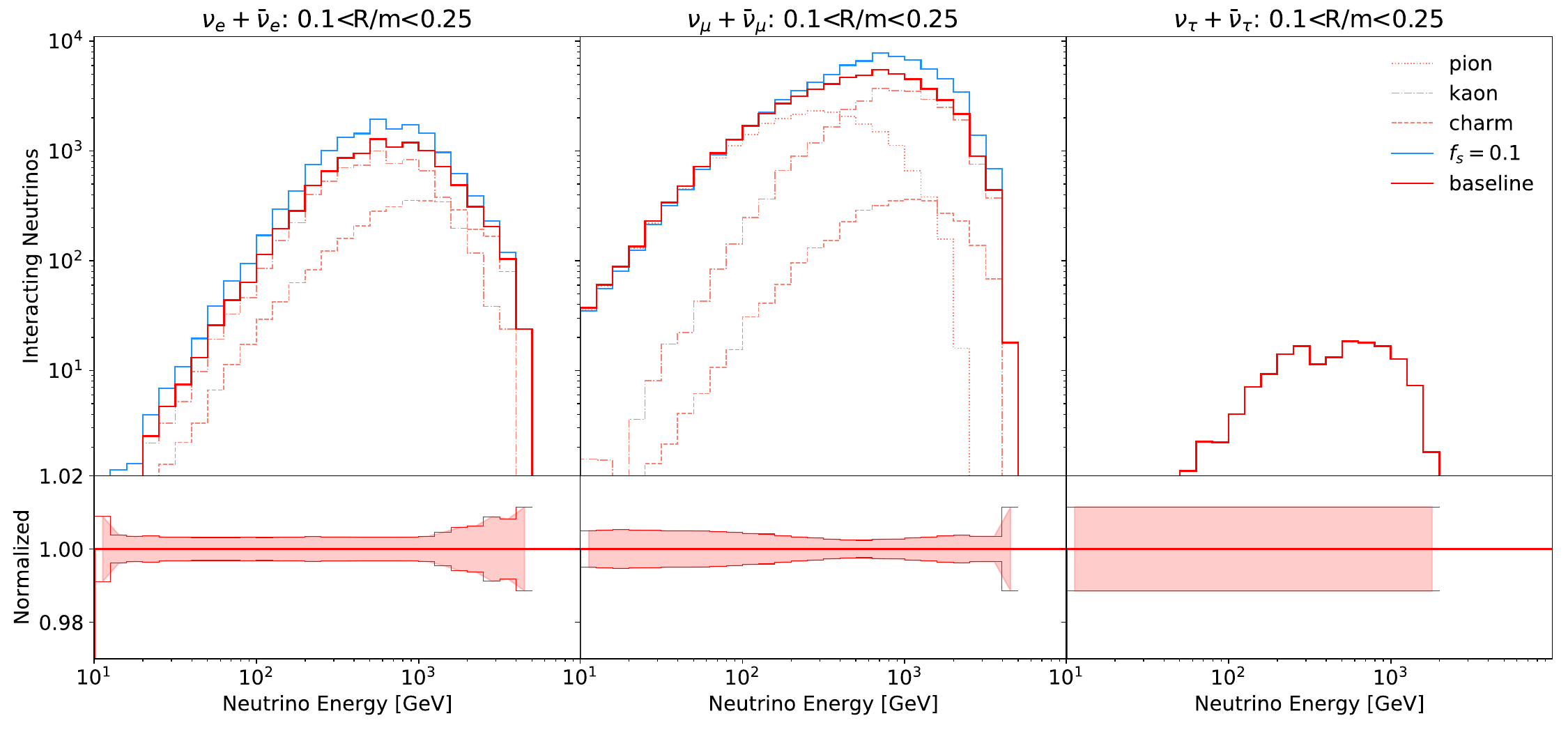}{0.8}
  \postscript{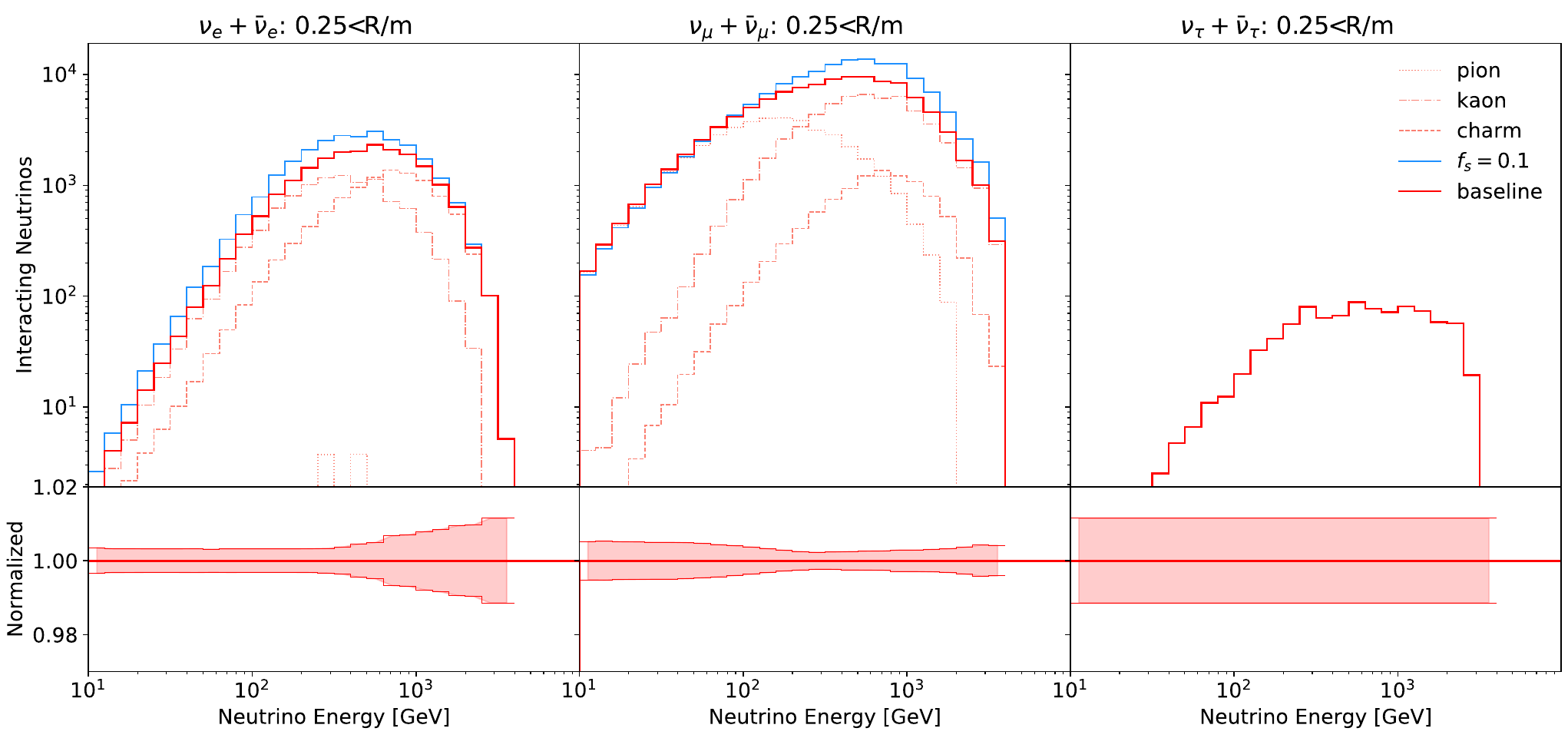}{0.79}
\caption{Neutrino energy spectra for electron neutrinos (left), muon neutrinos (middle), and tau neutrinos (right) passing through the ${\rm m}^2$ FLArE detector, assuming an integrated luminosity of $3~{\rm ab}^{-1}$, for three bins of radius $0<R/{\rm m} <0.1$, $0.1 <R/{\rm m}< 0.25$, and $R>0.25~{\rm m}$. \label{fig:flux}}
\end{figure}

Next, in line with our stated plan, we compare the results of the baseline model with predictions from  the one-parameter piKswap model with $f_s =0.1$. In Fig.~\ref{fig:flux}  we show the expected fluxes of electron neutrinos (left), muon neutrinos (middle) and tau neutrinos (right) at the FLArE detector. Results from the \textsc{Sibyll-2.3d} baseline model are shown as solid red lines and the predictions from piKswap model with $f_s=0.1$ as solid blue lines. The flux composition for the baseline model is also shown. The lower panels show the expected uncertainties of the flux model obtained using the hessian approach. We find that measurements of the neutrino fluxes at FLArE can probe $f_s$ at the sub-percent level. Therefore, we can conclude that these measurements will allow us to test the piKswap hypothesis for the origin of the UHECR muon puzzle by constraining strangeness production in the far-forward region

\acknowledgments
S.J.S and C.G.C. are partially supported by ANPCyT. L.A.A. is supported by the U.S. National Science
Foundation (NSF Grant PHY-2112527). F.K. is supported by the Deutsche Forschungsgemeinschaft under Germany's Excellence Strategy - EXC 2121 Quantum Universe - 390833306. J.F.S. is supported by Schmidt Futures, a philanthropic initiative founded by Eric and Wendy Schmidt, as part of the Virtual Institute for Astrophysics (VIA).

\begingroup
\bibliographystyle{ICRC}
\bibliography{references}
\endgroup

\end{document}